\documentclass[preprint]{article}

\usepackage{neurips_2025}

\usepackage[utf8]{inputenc} 
\usepackage[T1]{fontenc}    
\usepackage{hyperref}       
\usepackage{url}            
\usepackage{booktabs}       
\usepackage{amsfonts}       
\usepackage{nicefrac}       
\usepackage{microtype}      
\usepackage{xcolor}         

\usepackage{amsmath}
\usepackage{amssymb}
\usepackage{mathtools}
\usepackage{amsthm}

\theoremstyle{plain}

\theoremstyle{definition}

\theoremstyle{remark}

\usepackage[textsize=tiny]{todonotes}

\usepackage[dvipsnames]{xcolor}
\usepackage{pifont}
\definecolor{redx}{RGB}{180,0,0}
\definecolor{greenx}{RGB}{0,180,0}

\definecolor{redx}{RGB}{180,0,0}
\definecolor{greenx}{RGB}{0,180,0}
\usepackage{threeparttable}

\usepackage{caption}  
\usepackage{wasysym}

\usepackage[ruled,algo2e]{algorithm2e}
\usepackage{subcaption}
\usepackage[capitalise]{cleveref} 

\usepackage{enumitem}

\setlist[itemize]{leftmargin=*}

\usepackage[framemethod=TikZ]{mdframed}
\mdfdefinestyle{myframe}{%
    linecolor=gray!15!white,
    outerlinewidth=0.5pt,
    roundcorner=2pt,
    innertopmargin=2pt,
    innerbottommargin=2pt,
    innerrightmargin=2pt,
    innerleftmargin=2pt,
    backgroundcolor=gray!15!white
}

\usepackage[framemethod=TikZ]{mdframed}
\mdfdefinestyle{mydefframe}{%
    linecolor=cyan!10!white,
    outerlinewidth=0.5pt,
    roundcorner=10pt,
    innertopmargin=10pt,
    innerbottommargin=\baselineskip,
    innerrightmargin=10pt,
    innerleftmargin=10pt,
    backgroundcolor=cyan!10!white
}

\usepackage{comment}

\usepackage{pifont}

\usepackage{stmaryrd}
\newcommand{\enc}[1]{\llbracket #1 \rrbracket}

\usepackage{booktabs}
\usepackage{multirow}
\usepackage{tcolorbox}
\usepackage{booktabs}

\title{Fairly Compensated Distributed Information Retrieval and Augmentation for AI Agents}

\author{
  Yixiang Yao \quad Pasha Barahimi \quad Srivatsan Ravi \\[2pt]
  University of Southern California \\[2pt]
  \texttt{\{yixiangy, barahimi, srivatsr\}@usc.edu}
}

\begin{document}

\maketitle

\begin{abstract}
The increasing reliance of autonomous AI agents on external and distributed knowledge sources introduces a fundamental challenge for decentralized information marketplaces: retrieval agents must evaluate the quality and relevance of data before purchase, while data providers must avoid revealing valuable information prior to guaranteed compensation. This paradox becomes particularly critical in trustless multi-agent environments, where no centralized intermediary can enforce fairness between parties.
In this paper, we propose a fairly compensated protocol for distributed information retrieval and augmentation in autonomous agent networks. Our framework enables retrieval agents to securely evaluate and rank candidate documents without learning their plaintext contents, while ensuring that data providers are compensated only when valid information is successfully delivered. 
We further analyze the security properties of the protocol against malicious adversaries and evaluate its practical feasibility through implementations. Experimental results demonstrate that the proposed design is practical with current cryptographic infrastructures while preserving confidentiality, correctness, integrity, and fairness. We believe such mechanisms provide an important cryptographic foundation for trustworthy and economically sustainable decentralized knowledge marketplaces for future AI agent ecosystems.

\end{abstract}
\section{Introduction}
\label{sec:intro}

The emergence of Large Language Models (LLMs) \cite{openai_chatgpt,google_gemini,anthropic_claude} and autonomous AI agents \cite{hughes2025ai,durante2024agent} has revolutionized complex reasoning and decision-making tasks. However, the efficacy of these agents is strictly bounded by the quality and timeliness of their underlying knowledge base. As an example, Retrieval-Augmented Generation (RAG) \cite{arslan2024survey,cuconasu2024power,zhang2024raft} has surfaced as the standard architectural pattern to ground these models. More generically, in a multi-agent network, the data flow is multi-directional: an agent may query a database node, or two agents may engage in a peer-to-peer exchange where one possesses the specific domain expertise or environmental state required by the other. 
In any of these scenarios, the ``gold standard'' of data (often domain-specific, inter-organizational, or privacy-sensitive) is rarely centralized. Instead, it is distributed across a heterogeneous network of data providers, ranging from institutional repositories to specialized edge-agent nodes \cite{yao2026trustworthy}.

Despite the critical need for distributed data, current monetization models are fundamentally misaligned with data value. Existing frameworks typically rely on either rigid subscription models or volume-based pricing (e.g., pay-per-query), which fail to account for the intrinsic quality of the retrieved information. This creates a two-sided market failure. On the one side, from the aspect of AI agents, there is a pervasive junk-data risk, where agents incur significant costs for low-value or irrelevant responses. On the other side, for data providers, particularly smaller and specialized ones, the current model favors data monopolies because trusted ``big names'' naturally equal low risk. Without a mechanism to prove data quality prior to the transaction, high-quality but small providers remain undercompensated, causing them struggle to maintain the infrastructure necessary to compete.

\begin{figure}
    \centering
    \includegraphics[width=\linewidth]{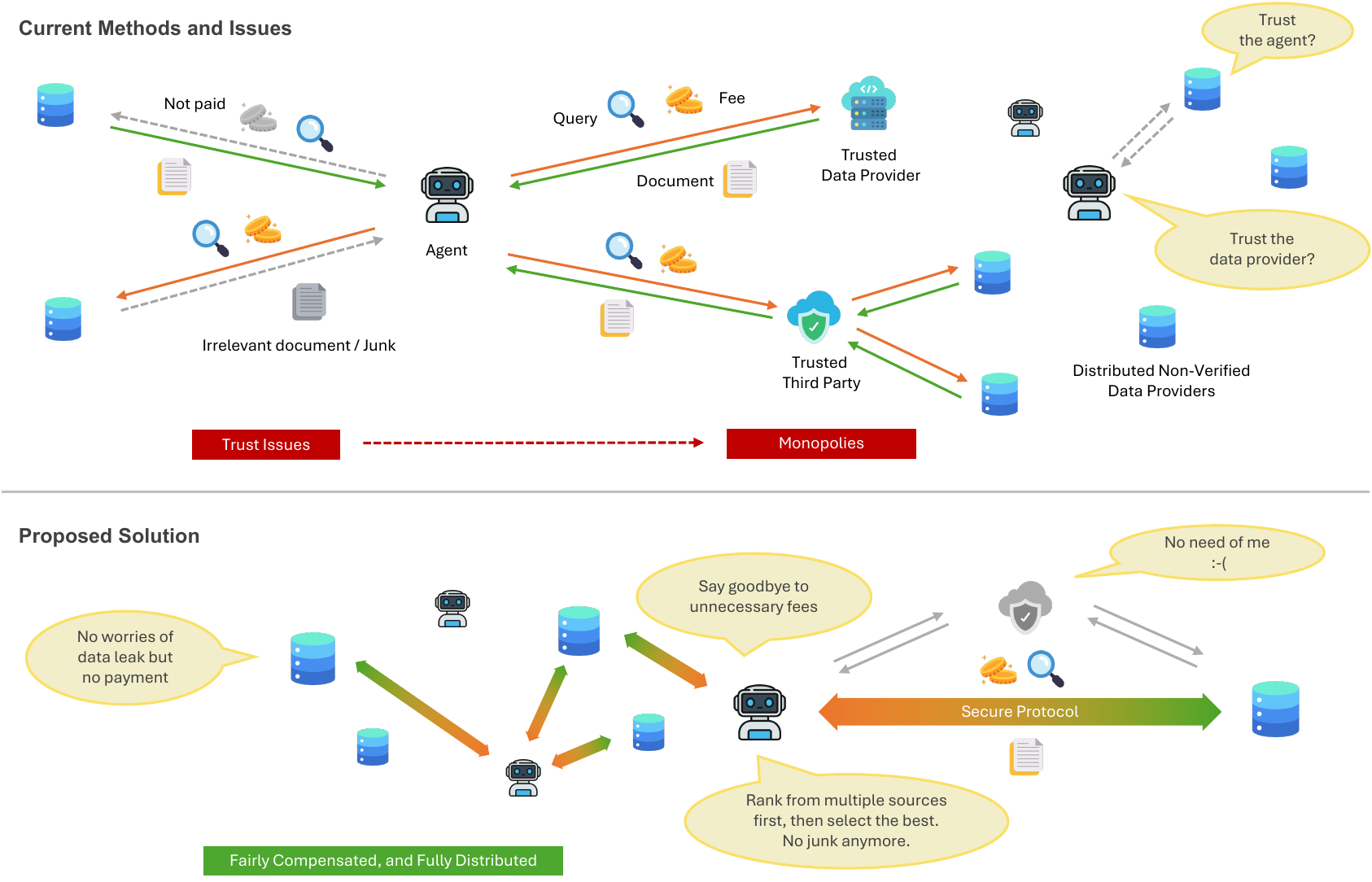}
    \caption{Illustration of current methods and the proposed solution.}
    \label{fig:intro}
\end{figure}

Ideally, in a fully distributed agent network with zero trust, the agent only compensates the provider who provides suitable data. However, it is challenging to archive with the three fundamental technical hurdles. First, the evaluation-disclosure paradox prevents transparent valuation: an agent must verify data utility before committing funds, yet a provider cannot reveal its data without confirming the payment. Second, an agent may need to poll multiple candidate nodes but lacks a secure mechanism to identify and compensate only the most relevant providers without leaking value to the others. Finally, executing these transactions requires an atomic settlement layer that guarantees ``delivery-versus-payment'' (DvP) in a peer-to-peer manner, ensuring that neither the data nor the payment is released without the other, all while operating without a centralized trusted intermediary.

In this paper, we propose a fairly compensated protocol specifically tailored for distributed information retrieval and augmentation in autonomous agent networks. Our framework resolves the evaluation-disclosure paradox by allowing agents to verify data against their specific query without compromising the provider’s underlying privacy. By enabling secure, pre-purchase comparisons across a distributed field of candidates, it empowers agents to select the most relevant information based on merit rather than brand reputation. Finally, the protocol facilitates the exchange of information and value as an atomic transaction to guarantee DVP without the involvement of a third party. By removing the requirement for prior trust or centralized intermediaries, our protocol establishes a privacy-preserving economic foundation for a truly decentralized knowledge marketplace.
\section{Preliminaries}
\label{sec:[preliminaries]}

\subsection{Problem Statement}


We formalize the problem as a multi-party protocol involving a set of $n$ \textit{distributed data providers} $\mathcal{S} = \{s_1, s_2, \dots, s_n\}$ and a set of $m$ \textit{autonomous agents} $\mathcal{C} = \{c_1, c_2, \dots, c_m\}$.

\paragraph{Retrieval}
An agent $c_i \in \mathcal{C}$ initiates a query $q \in \{0,1\}^*$ directed to a subset of providers $\mathcal{S}' \subseteq \mathcal{S}$. Each provider $s_j \in \mathcal{S}'$ identifies a candidate response $d_j$ from their local knowledge base $\mathcal{K}_j$:
\begin{equation}
    d_j = \text{Retrieve}(q, \mathcal{K}_j)
\end{equation}

Each $s_j$ transforms $d_j$ into a form $T(d_j)$ that hides the plaintext content from the agent $c_i$ but is able to evaluate with.

\paragraph{Qualification}
The agent $c_i$ can verify if $d_j$ is qualified for $q$ without learning its plaintext with the function $Q$, that is,
\begin{equation}
    \forall s_j \in \mathcal{S}', \quad c_i \text{ verifies } Q(T(d_j), q) \in \{0, 1\}.
\end{equation}


\paragraph{Settlement}
Based on the results, $c_i$ selects a subset of ``winning'' documents $\mathcal{D}^*$, and prepare a set of payments $\mathcal{P}^*$ to purchase. The transaction must execute an \textit{Atomic Information-Value Exchange} between the agent $c_i$ and each selected provider $s_j$. For a payment $p_j \in \mathcal{P}^*$ and document $d_j \in \mathcal{D}^*$, the final state transition must be:
\begin{equation}
    \text{State}_{\text{final}} = 
    \begin{cases} 
    (c_i \leftarrow d_j, s_j \leftarrow p_j) & \text{if both parties are honest} \\
    (c_i \leftarrow \perp, s_j \leftarrow \perp) & \text{if either party aborts}
    \end{cases}
\end{equation}
This exchange is conducted without a \textit{trusted third party}. This ensures that the agent cannot recover $d_j$ unless the payment $p_j$ is committed to the provider, and the provider cannot claim $p_j$ without revealing a $d_j$.

We say a transaction is \textit{fairly compensated} if the retrieval agent can verify the qualification and relevance of a document prior to purchase, while the data provider is guaranteed to receive compensation if and only if the correct qualified document is delivered. Consequently, providers supplying higher-quality or more relevant documents are more likely to be selected and compensated than those providing lower-quality responses.

\subsection{Homomorphic Encryption}


Homomorphic Encryption (HE) \cite{acar2018survey, fontaine2007survey} is a cryptographic primitive that enables computations to be performed directly on encrypted data without requiring decryption. Formally, an HE scheme for a plaintext space $\mathcal{M}$ and ciphertext space $\mathcal{C}$ allows the evaluation of an operator $\oplus_C$ on the underlying messages through a corresponding operator $\oplus_C$ on their ciphertexts. If $ct_1 = \text{Enc}(m_1)$ and $ct_2 = \text{Enc}(m_2)$, then:
\begin{equation}
    \text{Dec}(ct_1 \oplus_C ct_2) = m_1 \oplus_M m_2
\end{equation}
In concrete implementations, $\oplus_M$ typically represents fundamental arithmetic operations such as addition ($+$) or multiplication ($\times$), or logical operations such as and ($\wedge$), or ($\vee$) and not ($\neg$).

A typical HE scheme ${\epsilon}$ is defined by four operations:
\begin{itemize}
    \item $\mathsf{HE.KeyGen}_{\epsilon}(1^\lambda) \to (pk, sk)$: Given a security parameter $\lambda$, the algorithm generates a public key $pk$ for encryption and evaluation, a secret key $sk$ for decryption.
    \item $\mathsf{HE.Enc}_{\epsilon}(pk, m) \to ct$: Encrypts a plaintext message $m$ into a ciphertext $ct$ using the public key.
    \item $\mathsf{HE.Dec}_{\epsilon}(sk, ct) \to m$: Recovers the original message $m$ from the ciphertext using the secret key.
    \item $\mathsf{HE.Eval}_{\epsilon}(pk, f, \{ct_1, \dots, ct_n\}) \to ct_{\text{eval}}$: Takes the public key $pk$, a function $f$ (represented as a boolean or arithmetic circuit), and a set of ciphertexts, producing a ciphertext $ct_{\text{eval}}$ that encrypts the result of $f$ applied to the underlying plaintexts.
\end{itemize}

Modern HE research is primarily categorized by schemes optimized for specific data types. The BFV \cite{fan2012somewhat} and BGV \cite{brakerski2014leveled} schemes are optimized for exact integer arithmetic, and the CKKS scheme \cite{cheon2017homomorphic} is designed for approximate arithmetic on real or complex numbers. 

To further enhance security in distributed environments, \textit{Threshold Homomorphic Encryption (THE)} \cite{Desmedt2011,cramer2001multiparty} extends these schemes into a multi-party setting. In THE, the secret key $sk$ is not held by a single entity but is instead divided into $n$ shares $\{sk_1, sk_2, \dots, sk_n\}$ distributed among multiple participants. Decryption requires a threshold of parties ($t \leq n$) to collaborate in a partial decryption process. This ensures that no single (or less than $t$) provider or agent can unilaterally decrypt the data, providing a robust defense against colluding nodes and centralized points of failure in the distributed network. Specifically, for a scheme $\epsilon$, the distributed key generation and decryption are defined as below.

\begin{itemize}
    \item $\mathsf{THE.KeyGenLead}_{\epsilon}(1^{\lambda}, t, n) \to (pk_1, sk_1)$: The initial party $P_1$ generates a base key pair.
    \item $\mathsf{THE.KeyGenMain}_{\epsilon}(pk_{j-1}, \dots) \to (pk_{joint}, sk_j)$: Subsequent parties $P_j$ generate their own secret shares $sk_j$ and iteratively update the joint public key $pk_{joint}$.
    \item $\mathsf{THE.DecryptLead}_{\epsilon}(sk_1, ct) \to ct_{p1}$: The designated lead party initiates the decryption process by applying their secret share $sk_1$ to the ciphertext $ct$. This produces a partially decrypted ciphertext component $ct_{p1}$.
    \item $\mathsf{THE.DecryptMain}_{\epsilon}(sk_j,ct) \to ct_{pj}$: All other participating parties $P_j$ independently compute their partial decryption shares $ct_{pj}$ using their respective secret shares $sk_j$.
    \item $\mathsf{THE.DecryptFusion}_{\epsilon}(\{ct_{p1}, \dots, ct_{pt}\}) \to m$: The final aggregation step where the lead and main partial decryptions are combined. At least $t$ parties need to be involved to execute a correct decryption.
\end{itemize}


\subsection{Zero-Knowledge Proof}


Zero-Knowledge Proofs (ZKPs) \cite{fiege1987zero} are cryptographic protocols that enable a party, the \textit{prover} ($P$), to convince another party, the \textit{verifier} ($V$), that a specific statement is true without revealing any information beyond the validity of the statement itself. Formally, for a NP-language $\mathcal{L}$ associated with a polynomial-time decidable relation $\mathcal{R}$, a ZKP allows $P$ to prove that for a public input $x$, they possess a private witness $w$ such that $(x, w) \in \mathcal{R}$. 

A robust ZKP system must satisfy three fundamental properties:
\begin{itemize}
    \item Completeness: If $(x, w) \in \mathcal{R}$ and both the prover and verifier follow the protocol, the verifier will be convinced with overwhelming probability: $\Pr[\langle P(w), V \rangle(x) = \text{accept}] \geq 1 - \text{negl}(\lambda)$.
    \item Soundness: If $(x, w) \notin \mathcal{R}$, no malicious prover $P^*$ can convince the verifier that the statement is true, except with a negligible probability: $\Pr[\langle P^*, V \rangle(x) = \text{accept}] \leq \text{negl}(\lambda)$
    \item Zero-Knowledge: For every probabilistic polynomial-time verifier $V^*$, there exists a polynomial-time Simulator $\text{Sim}$ such that for all $(x, w) \in \mathcal{R}$: $\text{View}_{V^*}[P(w) \leftrightarrow V^*(x)] \approx \text{Sim}(x)$, where $\approx$ denotes computational indistinguishability.
\end{itemize}

In this work, we specifically leverage \textit{Zero-Knowledge Succinct Non-Interactive Arguments of Knowledge (zk-SNARKs)} \cite{bitansky2012extractable,parno2016pinocchio}. Unlike traditional interactive proofs, zk-SNARKs enable the prover to generate a single, non-interactive proof string $\pi$ that can be verified publicly. These systems are ``succinct'' because the proof size $|\pi|$ is typically $O(1)$ or $O(\text{poly} \log |w|)$, and the verification time is significantly faster than executing the underlying computation. 

\subsection{Blockchain and Smart Contracts}

A \textit{blockchain} \cite{zheng2018blockchain} is a decentralized, distributed ledger that maintains a continuously growing list of records, called blocks, which are linked and secured using cryptography. It provides a transparent and immutable infrastructure for data storage, ensuring that once a transaction is recorded and confirmed by network consensus, it cannot be altered or deleted. 

\textit{Smart contracts} \cite{nofer2017blockchain} are self-executing programs stored on the blockchain that automatically facilitate, verify, or enforce the performance of an agreement. By encoding the terms of a transaction directly into immutable code, these contracts move beyond simple data storage to provide programmable logic.

\textit{Ethereum} \cite{buterin2013ethereum} is a widely-used, decentralized, open-source blockchain platform which provides a Turing-complete execution environment known as Ethereum Virtual Machine (EVM), purpose-built for deploying and executing smart contracts. Ethereum's native currency, Ether (ETH), serves both as a medium of exchange and as ``gas'', the fee paid to compensate for the computational resources required to execute operations on the network. Smart contracts on Ethereum are implemented in \textit{Solidity} \cite{dannen2017introducing}, a statically typed, contract-oriented programming language that compiles to EVM bytecode.

A \textit{Hashed Timelock Contract (HTLC)} \cite{herlihy2018atomic} is a type of smart contract used in cryptocurrencies to enable secure, trustless, and time-bound transactions. It forces the recipient to acknowledge receipt of payment by a deadline using a cryptographic secret (``hashlock'') or else the funds are refunded to the sender (``timelock''). For example, Alice locks funds in a digital vault that Bob can only open by revealing a specific secret code before a countdown timer expires, or else the money returns to Alice.

\subsection{Related Works}

\paragraph{Retrieval-Augmented Generation and Autonomous Agents}

Retrieval-Augmented Generation (RAG) has emerged as a foundational paradigm for enhancing large language models with external and up-to-date knowledge. Early works such as RAG~\cite{lewis2020retrieval} combine parametric language models with non-parametric retrieval systems to improve factuality and knowledge coverage. More recently, autonomous agent frameworks including ReAct~\cite{yao2022react}, Toolformer~\cite{schick2023toolformer}, and AutoGen~\cite{wu2023autogenenablingnextgenllm} demonstrate that modern AI systems increasingly rely on continuous interactions with external tools, APIs, and distributed information sources. These developments indicate a broader transition from standalone language models toward interconnected agent ecosystems where information retrieval becomes a core operational primitive.

At the same time, recent works have recognized that retrieval itself incurs substantial computational and economic cost. Approaches such as Self-RAG~\cite{asai2024selfrag}, CRAG~\cite{yan2024crag}, Adaptive-RAG~\cite{jeong2024adaptiverag}, and FrugalRAG~\cite{java2025frugalrag} attempt to improve the efficiency and quality of retrieval by selectively invoking external knowledge sources or correcting low-quality retrieval outputs. These systems optimize the trade-off between retrieval cost and response quality, highlighting the growing importance of retrieval valuation in large-scale AI systems. However, existing RAG frameworks largely assume trusted and centralized retrieval infrastructures, and do not address how autonomous agents can securely evaluate and compensate distributed information providers in trustless environments.

\paragraph{Distributed AI and Decentralized Data Marketplaces}

The emergence of decentralized AI infrastructures further amplifies the need for trustworthy information exchange protocols. Platforms such as Ocean Protocol \footnote{https://oceanprotocol.com/} and Fetch.ai \footnote{https://www.fetch.ai/} aim to establish decentralized marketplaces for data, computation, and AI services. These systems introduce tokenized incentive mechanisms and enable economic interactions between distributed agents and providers. In parallel, multi-agent systems such as CAMEL~\cite{li2023camel} and Generative Agents~\cite{park2023generative} envision persistent autonomous agents that continuously exchange information and coordinate actions in open environments.


\paragraph{Fair Exchange and Trustless Settlement.}

The fair exchange problem has long been studied in distributed systems and cryptography, particularly in the context of exchanging digital assets between mutually untrusted parties. Recent works have further analyzed the practical security and incentive compatibility of HTLC-based systems. For example, He-HTLC~\cite{wadhwa2023hehtlc} revisits the incentive assumptions of classical HTLC protocols and proposes a more robust construction against strategic manipulation and bribery attacks.
These approaches primarily focus on secure payment settlement or fair delivery after the data has already been selected.

However, regardless of all these works, for the specific problem of fairly compensated distributed information retrieval for autonomous agents remains unexplored. In particular, prior research has not systematically investigated how retrieval agents can privately evaluate and compare candidate information from multiple distributed providers before purchase, while simultaneously ensuring that providers are compensated fairly and atomically only when valid information is successfully delivered. This gap becomes increasingly important as future AI ecosystems evolve toward decentralized and economically autonomous multi-agent environments.

\section{Method}
\label{sec:method}

In this section, we first present and examine the protocol in detail in \Cref{sec:protocol}, followed by a security analysis in \Cref{sec:security-analysis}.

\subsection{Protocol}
\label{sec:protocol}

Assume the retrieval agent is $P_0$, and there are two data providers $P_1$ and $P_2$. In practice, there could be more than two data providers.

\textbf{Phase 1: Ranking}

In Phase 1, $P_0$ issues a query to $P_1$ and $P_2$ to perform retrieval, and subsequently scores and ranks the returned documents without learning their plaintext contents. By the end, $P_0$ knows which document is worth the retrieval.

The protocol is illustrated in \Cref{fig:protocol-phase1}. In Step 1, $P_0$ initiates the distributed THE key generation protocol with $P_1$ and $P_2$. At the end of this step, each party holds its own private key, while $P_1$ and $P_2$ each establish a joint public key with $P_0$ (i.e., $pk_{joint1}$ and $pk_{joint2}$).

In Step 2, $P_0$ sends the query $q$ to both $P_1$ and $P_2$. In Step 3, $P_1$ and $P_2$ perform retrieval using $q$, obtain documents $d_1$ and $d_2$, encrypt them under their respective joint public keys, and return the ciphertexts $\enc{d_1}$ and $\enc{d_2}$ to $P_0$.

In Step 4, $P_0$ evaluates the encrypted documents using THE. It computes relevance scores via a homomorphic scoring function $S$, and simultaneously generates document hashes using $\mathsf{HashDoc}$ for subsequent integrity verification. The resulting encrypted scores and hashes are then sent back to $P_1$ and $P_2$.

In Step 5, $P_0$ performs partial decryption of the encrypted scores and hashes using its private key, and forwards the partially decrypted results to $P_1$ and $P_2$. In Step 6, $P_1$ and $P_2$ independently perform their own partial decryptions on the corresponding ciphertexts.

In Step 7, $P_1$ and $P_2$ combine their own partial decryption results (from Step 6) with those received from $P_0$ (from Step 5) to recover the plaintext scores and hashes. This step ensures that $P_0$ does not learn any sensitive information during the evaluation of $S$ and $\mathsf{HashDoc}$. If no information leakage is detected, $P_1$ and $P_2$ send their partial decryption shares to $P_0$.

In Step 8, $P_0$ combines its own partial decryption results (from Step 5) with those received from $P_1$ and $P_2$ (from Step 7) to obtain the final plaintext scores and hashes. Finally, in Step 9, $P_0$ compares the scores, selects the best document, and retains the corresponding hash value. If none of the retrieved documents are satisfactory (e.g., both scores are less than a minimum threshold), $P_0$ may terminate the protocol without proceeding to the subsequent phases.

\begin{figure}[t]
\centering
\begin{tcolorbox}[
    colback=black!10!white,
    colframe=black!70!white,
    boxrule=0.5pt,
    arc=1mm,
    left=1mm,
    right=1mm,
    top=1mm,
    bottom=1mm,
    width=\linewidth,
]
\renewcommand{\arraystretch}{1.2} 
\begin{enumerate}[leftmargin=*, label=\arabic*.]
    \item $P_0$ generates $(pk_0, sk_0) \gets \mathsf{THE.KeyGenLead}_{\epsilon}(1^{\lambda}, 2, 2)$. $t$ and $n$ are set to 2 to ensure decryption will be successful only when both $(P_0, P_1)$ or $(P_0, P_2)$ are involved. $P_0$ shares $pk_0$ with $P_1$ and $P_2$. $P_1$ and $P_2$ then generate $(pk_{joint1}, sk_1) \gets \mathsf{THE.KeyGenMain}_{\epsilon}(pk_0)$ and $(pk_{joint2}, sk_2) \gets \mathsf{THE.KeyGenMain}_{\epsilon}(pk_0)$, and share $pk_{joint1}$ and $pk_{joint2}$ to $P_0$ respectively.

    \item $P_0$ sends query $q$ to $P_1$ and $P_2$.

    \item $P_1$ and $P_2$ perform retrieval with $q$ and retrieve documents $d_1$ and $d_2$. They also encrypt both documents to $\enc{d_1} \gets \mathsf{HE.Enc}_{\epsilon}(pk_{joint1},d_1)$ and $\enc{d_2} \gets \mathsf{HE.Enc}_{\epsilon}(pk_{joint2},d_2)$. They send $\enc{d_1}$ and $\enc{d_2}$ to $P_0$.

    \item $P_0$ homomorphically executes a scorer $S$ on $\enc{d_1}$ and $\enc{d_2}$, $\enc{s_1} \gets \mathsf{THE.Eval}_{\epsilon}(pk_{joint1}, S, \enc{d_1})$ and $\enc{s_2} \gets \mathsf{HE.Eval}_{\epsilon}(pk_{joint2}, S, \enc{d_2})$, meanwhile homomorphically computes hashes $\enc{h_1} \gets \mathsf{THE.Eval}_{\epsilon}(pk_{joint1}, \mathsf{HashDoc}, \enc{d_1})$, $\enc{h_2} \gets \mathsf{THE.Eval}_{\epsilon}(pk_{joint2}, \mathsf{HashDoc}, \enc{d_2})$. $P_0$ sends $(\enc{s_1}, \enc{h_1})$ to $P_1$ and $(\enc{s_2}, \enc{h_2})$ to $P_2$.

    \item $P_0$ performs independent partial decryption on scores and hashes: $\enc{s_{1p0}} \gets \mathsf{THE.DecryptLead}_{\epsilon}(sk_0, \enc{s_1})$, $\enc{h_{1p0}} \gets \mathsf{THE.DecryptLead}_{\epsilon}(sk_0, \enc{h_1})$, $\enc{s_{2p0}} \gets \mathsf{THE.DecryptLead}_{\epsilon}(sk_0, \enc{s_2})$, $\enc{h_{2p0}} \gets \mathsf{THE.DecryptLead}_{\epsilon}(sk_0, \enc{h_2})$. $P_0$ sends the partially decrypted results $(\enc{s_{1p0}}, \enc{h_{1p0}})$ to $P_1$ and $(\enc{s_{2p0}}, \enc{h_{2p0}})$ to $P_2$.

    \item $P_1$ and $P_2$ perform independent partial decryption: $\enc{s_{1p1}} \gets \mathsf{THE.DecryptMain}_{\epsilon}(sk_1,\enc{s_1})$, $\enc{h_{1p1}} \gets \mathsf{THE.DecryptMain}_{\epsilon}(sk_1,\enc{h_1})$, $\enc{s_{2p2}} \gets \mathsf{THE.DecryptMain}_{\epsilon}(sk_2,\enc{s_2})$, $\enc{h_{2p2}} \gets \mathsf{THE.DecryptMain}_{\epsilon}(sk_2,\enc{h_2})$. $P_1$ and $P_2$ send their partially decrypted results to $P_0$.

    \item $P_1$ collects the partial decrypted results and fuses for the final decryption: $s_1 \gets \mathsf{THE.DecryptFusion}_{\epsilon}(\{\enc{s_{1p0}}, \enc{s_{1p1}}\})$, $h_1 \gets \mathsf{THE.DecryptFusion}_{\epsilon}(\{\enc{h_{1p0}}, \enc{h_{1p1}}\})$. $P_2$ can perform the similar operations to get decrypted results: $s_2 \gets \mathsf{THE.DecryptFusion}_{\epsilon}(\{\enc{s_{2p0}}, \enc{s_{2p2}}\})$, $h_2 \gets \mathsf{THE.DecryptFusion}_{\epsilon}(\{\enc{h_{2p0}}, \enc{h_{2p2}}\})$. If the decryption results $(s_1, h_1)$ and $(s_2, h_2)$ are with no information leak, $P_1$ and $P_2$ send their partial decrypted results $(\enc{s_{1p1}}, \enc{h_{1p1}})$ and $(\enc{s_{2p2}}, \enc{h_{2p2}})$ to $P_0$.

    \item $P_0$ collects the partial decrypted results and fuses for the final decryption: $s_1 \gets \mathsf{THE.DecryptFusion}_{\epsilon}(\{\enc{s_{1p0}}, \enc{s_{1p1}}\})$, $h_1 \gets \mathsf{THE.DecryptFusion}_{\epsilon}(\{\enc{h_{1p0}}, \enc{h_{1p1}}\})$, $s_2 \gets \mathsf{THE.DecryptFusion}_{\epsilon}(\{\enc{s_{2p0}}, \enc{s_{2p2}}\})$, $h_2 \gets \mathsf{THE.DecryptFusion}_{\epsilon}(\{\enc{h_{2p0}}, \enc{h_{2p2}}\})$.

    \item $P_0$ compares the score, and picks the best one. $P_0$ also keeps the corresponding hash value in plaintext (either $h_1$ or $h_2$) of that document.
\end{enumerate}
\end{tcolorbox}
\caption{Protocol - Ranking}
\label{fig:protocol-phase1}
\end{figure}


\textbf{Phase 2: Verification}

Assume that $d_1$ is selected as the preferred document (e.g., $s_1 > s_2$). Thus, $P_0$ tend to acquire $d_1$ from $P_1$ via an atomic swap protocol, ensuring fairness without requiring any trusted third party. Prior to the actual transaction in the next phase, a verification step is performed to guarantee the correctness and integrity of the exchanged data.

As illustrated in \Cref{fig:protocol-phase2}, this phase combines symmetric encryption with a zero-knowledge proof. In Step 1, $P_1$ samples a random symmetric key $k$ and encrypts the document $d_1$ to obtain the ciphertext $c_1$. It also computes a hash of the key as $h_k$. 

In Step 2, $P_1$ sends $(c_1, h_k)$ to $P_0$. At this point, $P_0$ possesses $c_1, h_k)$ along with $h_1$ obtained from Phase 1, where $h_1$ is the hash of the target document $d_1$. However, $P_0$ cannot yet verify whether the hidden key $k$ corresponding to $h_k$ correctly decrypts $c_1$ to the intended document $d_1$.

To address this, in Step 3, $P_1$ generates a zk-SNARK proof $\pi$ demonstrating knowledge of a key $k$ such that $k$ correctly decrypts $c_1$ to the document $d_1$ whose hash matches the value obtained by $P_0$ in Phase 1.

In Step 4, $P_1$ sends $\pi$ to $P_0$, who verifies the proof. If the verification succeeds, $P_0$ is assured that the ciphertext $c_1$ corresponds to the correct document and $k$ will properly decrypt it. This guarantees the correctness of the forthcoming transaction.

\begin{figure}[t]
\centering
\begin{tcolorbox}[
    colback=black!10!white,
    colframe=black!70!white,
    boxrule=0.5pt,
    arc=1mm,
    left=1mm,
    right=1mm,
    top=1mm,
    bottom=1mm,
    width=\linewidth,
]
\renewcommand{\arraystretch}{1.2}
\begin{enumerate}[leftmargin=*, label=\arabic*.]
    \item With a symmetric encryption algorithm $\mathsf{SYM}$, $P_1$ samples a random symmetric key $k$ and encrypts $d_1$ to obtain $c_1 = \mathsf{SYM.Enc}_k(d_1)$. It also computes $h_k = \mathsf{HashKey}(k)$.
    \item $P_1$ sends $c_1, h_k$ to $P_0$. 
    \item $P_1$ generates a zk-SNARK proof $\pi$:
    \begin{equation}
        \text{I know $k$ such that } \\
        \text{(i) }\mathsf{HashKey}(k)=h_k \\
        \text{, and } \\
        \text{(ii) }\mathsf{HashDoc}(\mathsf{SYM.Dec}_k(c_1))=h_1
    \end{equation}
    \item $P_1$ sends $\pi$ to $P_0$. $P_0$ verifies this proof. If valid, $P_0$ is mathematically certain that the key hidden behind $h_k$ is the correct one.
\end{enumerate}
\end{tcolorbox}
\caption{Protocol - Verification}
\label{fig:protocol-phase2}
\end{figure}

\textbf{Phase 3: Transaction}

At this stage, $P_0$ proceeds to pay $P_1$ in exchange for the selected document $d_1$. The transaction is carried out over a decentralized network without relying on any trusted third party. As illustrated in \Cref{fig:protocol-phase3}, this is achieved using a smart contract, specifically a Hash Time-Locked Contract (HTLC), to ensure fairness and atomicity.

In Step 1, $P_0$ locks the payment into a smart contract with the condition that the funds can only be claimed by revealing $k$. Additionally, a timeout parameter $t$ is set, after which the funds are refunded to $P_0$ if unclaimed.

In Step 2, upon observing the locked funds on the blockchain, $P_1$ submits the key $k$ to the smart contract in order to redeem the payment. Once this transaction is confirmed, $k$ is publicly recorded on the ledger.

In Step 3, $P_0$ monitors the blockchain, retrieves $k$, and uses it to decrypt the ciphertext locally, recovering the plaintext document $d_1$.

\begin{figure}[t]
\centering
\begin{tcolorbox}[
    colback=black!10!white,
    colframe=black!70!white,
    boxrule=0.5pt,
    arc=1mm,
    left=1mm,
    right=1mm,
    top=1mm,
    bottom=1mm,
    width=\linewidth,
]
\renewcommand{\arraystretch}{1.2}
\begin{enumerate}[leftmargin=*, label=\arabic*.]
    \item $P_0$ deposits the compensation funds into a smart contract with the condition: "These funds can be claimed by anyone who provides a preimage $k$ such that $\mathsf{HashKey}(k)=h_k$." (Optionally, if the deposit is not claimed within time $t$, $P_0$ gets refunded.")
    \item $P_1$ sees the funds are locked on the blockchain. To take the money, $P_1$ must submit $k$ to the blockchain as the input to the transaction. Once the transaction is mined, $k$ becomes public on the ledger.
    \item $P_0$ observes the blockchain, reads $k$, and uses it to decrypt $c_1$ locally: $d_1=\mathsf{SYM.Dec}_k(c_1)$.
\end{enumerate}
\end{tcolorbox}
\caption{Protocol - Transaction}
\label{fig:protocol-phase3}
\end{figure}

\subsection{Security Analysis}
\label{sec:security-analysis}

\subsubsection{Threat Model}
\label{sec:threat-model}

We consider a \emph{malicious adversary model}, where parties may arbitrarily deviate from the prescribed protocol in an attempt to gain advantage, rather than merely following the protocol and trying to infer additional information. In particular, adversaries may generate malformed messages, provide inconsistent inputs, or strategically abort the protocol.

We consider two types of adversaries: (i) a malicious retrieval agent $P_0$, modeled as $\mathcal{A}_a$, and (ii) malicious data providers $P_1$ or $P_2$, modeled as $\mathcal{A}_d$.

The adversary $\mathcal{A}_a$ aims to obtain the plaintext content of documents (e.g., $d_1$ or $d_2$) without completing the payment, potentially by deviating from the protocol or exploiting intermediate computation results. Conversely, $\mathcal{A}_d$ aims to receive payment without providing the correct document, for example by supplying malformed ciphertexts, incorrect decryption keys, or inconsistent data.

We assume that there is no collusion between parties. Furthermore, all underlying cryptographic primitives (THE, symmetric encryption, hash functions, and zk-SNARKs) are assumed to be secure, and the blockchain correctly enforces the semantics of HTLCs.

\subsubsection{Security Definition}
\label{sec:security}

We say the protocol is \emph{secure} if it satisfies the following properties against any probabilistic polynomial-time adversary:

\begin{itemize}
    \item \textbf{Confidentiality:} No adversary $\mathcal{A}_a$ can learn any information about the plaintext document beyond what is revealed by the final output, unless the payment is completed.
    
    \item \textbf{Correctness:} If the protocol completes successfully, the document obtained by $P_0$ is exactly the document that was evaluated and selected in Phase~1.
    
    \item \textbf{Fairness:} Either (i) $P_0$ obtains the correct document and the data provider receives payment, or (ii) neither party gains advantage (i.e., no document is revealed and the payment is refunded).
    
    \item \textbf{Integrity:} The document delivered in the transaction phase must match the document committed during the ranking phase.
\end{itemize}

\subsubsection{Security Analysis}

We analyze the security of the protocol against both adversaries across the three phases.

\paragraph{Security against $\mathcal{A}_a$ (malicious $P_0$).}

\begin{itemize}
    \item \textbf{Phase 1.}  All documents are protected under threshold homomorphic encryption as ciphertexts $\enc{d_1}$ and $\enc{d_2}$. Due to the semantic security of THE, $\mathcal{A}_a$ cannot learn any information about the plaintext documents. Moreover, although $P_0$ evaluates the scoring function $S$ and $\mathsf{HashDoc}$, the results remain encrypted and require cooperation from $P_1$ or $P_2$ for decryption. The threshold structure ensures that $\mathcal{A}_a$ alone cannot recover any partial or full plaintext values, including scores or hashes.

    \item \textbf{Phase 2.}  The symmetric key $k$ is never revealed to $P_0$ during this phase. Instead, $P_1$ provides a zk-SNARK proof $\pi$ that attests to the correctness of $k$ without revealing it. By the zero-knowledge property, $\mathcal{A}_a$ learns nothing about $k$ beyond its validity. Therefore, $\mathcal{A}_a$ cannot decrypt $c_1 = \mathsf{SYM.Enc}_k(d_1)$.

    \item \textbf{Phase 3.}  To obtain $k$, $\mathcal{A}_a$ must trigger the HTLC by locking sufficient funds. Without depositing the payment, $\mathcal{A}_a$ cannot induce $P_1$ to reveal $k$. Hence, the protocol enforces fairness: $P_0$ obtains the decryption key if and only if payment is made.
\end{itemize}

\paragraph{Security against $\mathcal{A}_d$ (malicious $P_1$ or $P_2$).}

\begin{itemize}
    \item \textbf{Phase 1.}  The document hash $h_1$ (or $h_2$) is computed by $P_0$ via homomorphic evaluation, ensuring that it is bound to the encrypted document submitted by the data provider. This prevents $\mathcal{A}_d$ from later substituting a different document without detection.

    \item \textbf{Phase 2.}  If $\mathcal{A}_d$ attempts to provide an incorrect ciphertext $c_1'$ or a key $k'$ that does not correspond to the intended document, the zk-SNARK proof will fail. In particular, the proof enforces that:
\[
\mathsf{HashDoc}(\mathsf{SYM.Dec}_{k}(c_1)) = h_1,
\]
which binds the encrypted content to the previously agreed hash. Therefore, $\mathcal{A}_d$ cannot convince $P_0$ of the validity of an incorrect document.

    \item \textbf{Phase 3.}  The HTLC requires revealing the preimage $k$ satisfying $\mathsf{HashKey}(k) = h_k$ in order to claim the payment. Since $k$ has already been validated in Phase 2, $\mathcal{A}_d$ must reveal the correct key to obtain the funds. Upon revelation, $P_0$ can immediately decrypt $c_1$ to recover $d_1$. If $\mathcal{A}_d$ refuses to reveal $k$, the timeout mechanism ensures that the funds are refunded to $P_0$.
\end{itemize}

The protocol satisfies the security definition in \Cref{sec:security} against malicious adversaries defined in \Cref{sec:threat-model}. Confidentiality holds because documents remain encrypted under THE and symmetric encryption, and keys are protected by zero-knowledge proofs until payment is made. Correctness and integrity follow from the binding between ciphertexts and hashes enforced by $\mathsf{HashDoc}$ and the zk-SNARK proof, ensuring that the delivered document matches the one evaluated in Phase~1. Fairness is guaranteed by the HTLC mechanism, which enforces atomic exchange: the data provider can only obtain payment by revealing the valid key, and otherwise the payment is refunded.

\section{Practical Challenges and Implementation Choices}

While the proposed protocol provides strong security and fairness guarantees at the theoretical level, translating these cryptographic components into a practical system introduces several non-trivial engineering and performance challenges. In this section, we discuss the major implementation choices and practical trade-offs encountered in realizing the protocol.

\subsection{HE schemes}
THE provides flexibility for privacy-preserving computation, but the choice of the underlying homomorphic encryption scheme significantly affects both functionality and performance. Different HE schemes are optimized for different types of operations. For example, BGV is well suited for exact integer arithmetic, CKKS is designed for approximate floating-point computation, while FHEW/TFHE \cite{ducas2015fhew,chillotti2020tfhe} are optimized for Boolean and logical operations.

However, using a single scheme to efficiently support all functionalities required by the protocol is non-trivial. Schemes such as FHEW can theoretically realize arbitrary computations through Boolean circuits, but they are inefficient for arithmetic-heavy workloads because additions and multiplications must be decomposed into large numbers of gate-level operations. In contrast, BGV and CKKS natively support arithmetic operations efficiently, but they do not directly support logical operations or comparisons.

Modern neural retrieval or language models typically require non-linear activation functions, which are difficult to evaluate directly under HE. In CKKS-based systems, these functions are usually approximated with low-degree polynomials to reduce multiplicative depth and improve efficiency. Although this makes CKKS attractive for machine learning workloads, it introduces approximation errors and complicates compatibility with cryptographic primitives that rely on exact modular arithmetic.

One possible solution is to combine multiple schemes and perform scheme switching during computation. However, scheme switching introduces substantial computational overhead and implementation complexity. An alternative approach is to run different schemes in parallel, which increases memory consumption but avoids repeated switching costs.

In our implementation, we prioritize simplicity and compatibility across all protocol phases by using a single scheme throughout the system. Although CKKS would naturally fit similarity scoring due to its support for floating-point arithmetic, it is not well suited for implementing the Poseidon hash function, which relies heavily on exact modular arithmetic over finite fields. Consequently, we adopt the BGV scheme, which supports modular arithmetic directly and integrates more naturally with both Poseidon hashing and threshold decryption.

It is important to note that THE is not universally supported for all HE schemes. In practice, threshold variants are available only for selected schemes in existing libraries. For example, some libraries do not provide threshold implementations for FHEW/TFHE.

\subsection{Scoring}

The scoring function used in Phase~1 can, in principle, be any function implementable under homomorphic encryption. Existing research has demonstrated HE-compatible adaptations of modern machine learning models, including neural retrieval systems and transformer-based language models. However, these approaches often require sophisticated approximation techniques and incur substantial computational overhead.

In this work, our primary objective is to evaluate the feasibility and system overhead of the proposed protocol rather than optimizing retrieval quality. Therefore, we adopt a lightweight scoring mechanism based on the dot product between vectors. The dot-product operation is directly supported by the BGV scheme and can be evaluated efficiently with low multiplicative depth. This choice enables us to isolate the cryptographic overhead introduced by THE and Poseidon hashing without introducing additional complexity from large neural models. More advanced scoring models can be integrated into the framework in future work as HE implementations continue to improve.

\subsection{Hashing}

The role of hashing in the protocol is to ensure document integrity across different phases. In particular, the hash value computed during the ranking phase binds the selected ciphertext to the final document delivered in the transaction phase, preventing malicious providers from replacing or modifying the content after evaluation.

Although SHA-family \cite{sha256} hash functions are widely used in conventional systems, they are not ideal for zero-knowledge proof systems because they are optimized for bitwise operations, which translate into expensive arithmetic circuits. In contrast, Poseidon \cite{grassi2021poseidon, grassi2023poseidon2} is specifically designed for finite-field arithmetic, where its internal permutation consists mainly of modular additions, multiplications, and exponentiations. Compared to traditional hash functions such as SHA-family, Poseidon requires significantly fewer arithmetic constraints in zk-SNARK circuits, making proof generation and verification substantially more efficient for ZKP systems.

Theoretically, Poseidon should also integrate naturally with HE because its internal operations are arithmetic-based rather than Boolean-based. Nevertheless, implementing Poseidon efficiently under HE introduces practical challenges. In many HE implementations, the maximum supported ciphertext modulus is limited bits. 
However, Poseidon typically relies on large prime fields to achieve the desired security level.

To address this limitation, we employ the Chinese Remainder Theorem (CRT) \cite{ding1996chinese} technique. CRT decomposes computations over a large modulus into several smaller moduli that fit within the supported ciphertext space. In our implementation, large inputs are partitioned into multiple CRT components, each represented with a relatively small but HE compatible modulus. Although CRT increases the number of ciphertexts and homomorphic operations, it enables practical Poseidon evaluation under current HE constraints while preserving compatibility with the zk-SNARK verification phase.

\subsection{Symmetric Encryption}

A natural approach for delivering document to $P_0$ would be to reuse the keys and ciphers of THE from Phase 1. For example, the data provider could partially decrypt the encrypted document and allow $P_0$ to complete the decryption collaboratively. However, such a design is impractical for two main reasons:
First, public-key cryptographic systems, including HE schemes, incur significantly higher computational overhead than symmetric encryption. Using HE for full document transfer would therefore introduce unnecessary performance costs, especially for large documents.
Second, implementing homomorphic encryption primitives directly inside zero-knowledge proof systems is highly inefficient and lacks mature tooling support. Existing zk-SNARK frameworks are much better optimized for symmetric cryptographic primitives than for lattice-based HE operations. Reproducing the correctness of HE decryption inside a ZKP circuit would substantially increase circuit complexity and proving time.

For these reasons, we separate the confidential computation phase from the document transfer phase. After ranking and verification, the selected document is encrypted using a standard symmetric encryption scheme with a randomly generated key $k$. The correctness of this encryption is then verified through a zk-SNARK proof, while the key itself is released fairly through the HTLC mechanism.
This design allows the protocol to leverage the strengths of both cryptographic paradigms: HE enables privacy-preserving ranking and evaluation, while symmetric encryption provides efficient large-scale data transfer and seamless integration with zero-knowledge proofs.

\subsection{Transaction}


From a practical standpoint, the core problem this work addresses is the fair exchange of digital goods for payment: a client wishes to purchase an encrypted document from a provider, but neither party is willing to act first without a guarantee of reciprocation. If the client pays first, the provider may never deliver the decryption key. If the provider reveals the key first, the client may refuse to pay. This is a classical instance of the fair exchange problem, and traditional solutions require a trusted third party to act as an escrow, introducing a single point of failure and a trust assumption that is undesirable in a decentralized setting.

The HTLC resolves this by replacing the trusted third party with cryptographic guarantees enforced directly by the blockchain. The client locks the agreed payment in the HTLC against the hashlock $h = \text{Poseidon}(K)$, using the \texttt{createLock} function. The provider is then incentivized to call \texttt{withdraw} and reveal $K$ on-chain to collect the payment. The moment \texttt{withdraw} is executed, $K$ becomes publicly visible in the transaction data, allowing the client to retrieve it and decrypt the document. The two actions, payment release and key revelation, are thus collapsed into a single atomic transaction, making it impossible for one to occur without the other. If the provider fails to act before the timelock expires, the client reclaims their funds via \texttt{refund}, suffering no financial loss.

The \texttt{minLockDuration} parameter defines the minimum amount of time that must elapse between the creation of a lock and its expiration. This lower bound ensures that the provider has sufficient time to submit the \texttt{withdraw} transaction after the lock is created, accounting for network latency and block inclusion delays. Without this constraint, a malicious client could create a lock with an expiration set only seconds in the future, making it practically impossible for the provider 
to reveal the preimage and collect payment before the timelock expires, effectively forcing a refund despite the provider having acted in good faith. The appropriate value for \texttt{minLockDuration} depends on the target network and its expected congestion, on a live network such as Ethereum mainnet, it should be set to a value large enough to comfortably accommodate worst-case block inclusion times.

Despite the conceptual simplicity of the HTLC protocol, several practical challenges arise during implementation on Ethereum. The most notable is that the recipient does not receive the exact amount upon withdrawal. Since the \texttt{withdraw} transaction must be submitted and executed on-chain, the gas cost is deducted from the caller's account, meaning the recipient receives the full locked amount in ETH, but their net balance increased is slightly less than expected due to the gas fee paid to execute the transaction. This is an inherent property of the EVM execution model and not a flaw in the contract logic.

A second challenge is the time-sensitivity of the protocol: the recipient must submit the \texttt{withdraw} transaction before the lock expires, which can be problematic during periods of network congestion when transactions may be delayed or gas prices may spike, potentially forcing the recipient to pay a higher fee to ensure timely inclusion.

Finally, the deployment of the Poseidon hash function itself presents a practical challenge, as Poseidon is not a native EVM opcode. Its Solidity implementation must be deployed as a separate library and linked to the contract at deployment time, adding complexity to the deployment process compared to contracts that rely solely on built-in hash functions such as SHA-256.

\section{Evaluation}
\label{sec:evaluation}

\subsection{Settings}

The primary objective of the evaluation is to analyze the performance characteristics and computational overhead of the proposed protocol. To achieve a controlled and reproducible experimental environment, we use randomly generated data as protocol inputs. This approach enables systematic testing across different configurations while avoiding biases introduced by application-specific datasets.

In Phase 1, THE is implemented using the Python interface of OpenFHE \cite{al2022openfhe}, which provides Python bindings for the core cryptographic functionalities of the OpenFHE C++ library. The protocol is implemented based on the BGV homomorphic encryption scheme operating in multi-party threshold mode.

In Phase 2, the zero-knowledge proof (ZKP) component is implemented using Circom \cite{belles2022circom}, a domain-specific language designed for constructing arithmetic circuits for zk-SNARK systems. Circom includes a compiler that generates Rank-1 Constraint System (R1CS) representations together with the corresponding circuit constraints and witness generation programs. For proof generation and verification, we adopt the Groth16 \cite{cryptoeprint:2016/260} zk-SNARK protocol due to its efficiency in proof size and verification cost.

In Phase 3, we evaluate the proposed hash time-locked contract (HTLC) implementation based on the Poseidon hash function. The benchmark results are collected over 20 independent runs on both a local Hardhat network \footnote{https://www.hardhat.org/} and the Sepolia Ethereum testnet \footnote{https://github.com/eth-clients/sepolia}. The local Hardhat environment is used to isolate and measure the computational overhead of smart contract execution under controlled conditions, eliminating the influence of network latency, block production intervals, and congestion. In contrast, the Sepolia testnet provides performance measurements under realistic deployment conditions on a live proof-of-stake Ethereum network.

Unless otherwise specified, all experiments are conducted under the minimal deployment setting consisting of two data providers ($P_1$ and $P_2$) and one retrieval agent ($P_0$). Extending the protocol to additional participants does not introduce fundamental changes to the protocol workflow; instead, the resulting computational and communication overhead scales approximately linearly with the number of participating agents.


\subsection{Phase 1}

In the first phase of the protocol, we evaluate the impact of input size on the overall execution time. The input length ranges from 2 bytes to 32 bytes. For inputs smaller than the supported ciphertext capacity, the plaintext is padded to the nearest feasible size before encryption and evaluation.

Due to the limitation that the homomorphic encryption scheme supports ciphertext moduli of at most 60 bits in OpenFHE, only the 2-byte and 4-byte inputs can be processed directly using a native Poseidon hash implementation under homomorphic encryption (HE). For larger inputs, namely 8B, 16B, and 32B, we adopt a CRT-based encoding strategy. In these settings, each CRT modulus is configured with a 32-bit decomposition size, resulting in CRT counts of 2, 4, and 8, respectively.

To better understand the computational overhead introduced at each stage of the protocol, we measure the execution time of every individual step. As shown in \Cref{tab:p1_time_costs}, the Poseidon hash evaluation dominates the total runtime across all configurations. For smaller inputs (2B and 4B), the hash computation cost remains relatively close because both cases are processed directly without CRT decomposition. However, when the input size increases and CRT encoding is introduced, the evaluation time grows significantly. Despite this increase, the runtime scales approximately linearly with respect to both the input length and the number of CRT components.

This observation also motivates the choice of a 32-bit CRT basis. Compared with a 16-bit CRT configuration, the 32-bit setting introduces only a marginal increase in computational overhead while reducing the required CRT count by half, thereby improving overall efficiency. Similar scaling behavior can also be observed in the key generation and encryption stages.

For the scoring stage, we evaluate only a lightweight dot-product operation. Consequently, the computation time for scoring remains negligible across all input sizes. Likewise, the decryption phase, including both partial decryption and fusion operations, incurs only minimal overhead and remains consistently fast throughout the experiments.

\begin{table}[t]
\centering
\caption{Execution time of each step in Phase 1 of the protocol, measured in seconds. The table reports the runtime overhead for different input sizes ranging from 2B to 32B. Larger inputs are processed using CRT decomposition, which increases the number of ciphertext components and consequently the homomorphic evaluation cost. Poseidon hash evaluation is the dominant bottleneck, while scoring and decryption-related operations remain lightweight across all configurations.}
\label{tab:p1_time_costs}
\begin{tabular}{lccccc}
\toprule
\textbf{Steps} & \textbf{2B} & \textbf{4B} & \textbf{8B} & \textbf{16B} & \textbf{32B} \\
\midrule
Key generation & 1.49 & 1.52 & 4.16 & 8.01 & 15.92 \\
Encryption & 0.22 & 0.23 & 0.81 & 1.71 & 3.27 \\
Evaluation - Scorer & 0.02 & 0.03 & 0.04 & 0.04 & 0.04 \\
Evaluation - Hash & 6.59 & 7.10 & 22.10 & 44.96 & 86.72 \\
$P_0$'s partial decrypt & 0.04 & 0.05 & 0.05 & 0.05 & 0.11 \\
$P_1/P_2$'s partial decrypt & 0.04 & 0.05 & 0.05 & 0.05 & 0.10 \\
$P_0$'s fusion & 0.01 & 0.01 & 0.01 & 0.01 & 0.01 \\
$P_1/P_2$'s fusion & 0.01 & 0.01 & 0.01 & 0.01 & 0.01 \\
\bottomrule
\end{tabular}
\end{table}

\subsection{Phase 2}



\begin{table}[h]
\centering
\caption{Relations between input sizes vs Power of Tau.}
\label{tab:phase2-tau}
\begin{tabular}{l|rrrrr}
\toprule
\textbf{AES input size (byte)} & 16 & 32 & 48 & 64 & 80 \\
\midrule
\textbf{p ($2^p$)} & 16 & 16 & 17 & 17 & 18 \\
\textbf{ptau file size (MB)} & 73 & 73 & 145 & 145 & 289 \\
\textbf{Non-linear constraints} & 27856 & 45632 & 63408 & 81184 & 98960 \\
\textbf{Linear constraints} & 14275 & 19334 & 24393 & 29452 & 34511 \\
\textbf{Total constraints} & 42131 & 64996 & 87801 & 110636 & 133471 \\
\textbf{Public inputs (bits)} & 130 & 258 & 386 & 514 & 642 \\
\textbf{Private inputs (bits)} & 256 & 256 & 256 & 256 & 256 \\
\bottomrule
\end{tabular}
\end{table}

The Phase~2 verification circuit is implemented in Circom and instantiated using the Groth16 zk-SNARK protocol. The circuit takes four logical inputs: three public inputs and one private input. The public inputs consist of the hash of the symmetric key ($\mathsf{HashKey}(k)$), the hash of the document ($\mathsf{HashDoc}(d_1)$), and the AES ciphertext $c_1$ of the document. The private input is the symmetric key $k$. Note that the ciphertext is encoded as multiple field elements, which can be adjusted according to the document size.

Groth16 requires a trusted setup consisting of two phases. The first phase, known as the \emph{Powers of Tau}, is circuit-independent and can be reused across different circuits. The output is a \textit{ptau} file. The second phase is circuit-specific and depends on the compiled constraint system, i.e., R1CS in our settings.

A key parameter in the setup is $2^p$, which determines the maximum number of constraints supported in the circuit. For correctness, the total number of constraints in the circuit must be strictly less than $2^p$. As shown in \Cref{tab:phase2-tau}, larger document sizes lead to larger circuits, requiring higher values of $p$.

As the AES input size increases, the number of constraints grows approximately linearly. This is expected, as both AES decryption and hash computation are applied over the entire input. Consequently, the required Power of Tau parameter increases stepwise (e.g., from $2^{16}$ to $2^{18}$), which in turn increases the size of the ptau file and proving keys.

The most computationally expensive component is the generation of the ptau file. While this step takes only seconds for small $p$, it can take several hours for larger values. However, this cost is incurred only once and can be amortized across multiple circuits and deployments. In contrast, the circuit-specific Phase~2 setup and proof generation both complete within seconds in our experiments. The time cost of the verification of the proof is also negligible.

The size of the private input remains constant across all configurations, as the AES key has a fixed length. In contrast, the number of public inputs scales with the ciphertext size, since the encrypted document is provided as part of the statement.
Due to the use of AES in CTR mode within Circom, the input size must be a multiple of the block size (16 bytes). Documents that do not align with this requirement are padded to the nearest multiple.

\subsection{Phase 3}
To better quantify the practical overhead introduced by adopting Poseidon as the hash function, a comparison is drawn against a reference SHA-256-based HTLC implementation\footnote{https://github.com/chatch/hashed-timelock-contract-ethereum}. 
Since Poseidon is not a native EVM opcode and must be deployed as an external library
\footnote{https://www.npmjs.com/package/poseidon-solidity}
, while SHA-256\cite{sha256} is directly supported as a precompile, this comparison establishes whether the ZK-friendliness of Poseidon comes at a meaningful cost in terms of gas consumption and execution latency, or whether it can be adopted without significant practical penalty.

\paragraph{Gas Consumption}
Gas is the unit of measurement for the computational work required to execute operations on the Ethereum network. Every instruction executed by the EVM has an associated gas cost, and the total gas consumed by a transaction reflects the sum of all operations performed during its execution. Users specify a gas price, denominated in gwei, where $1$ gwei $= 10^{-9}$ ETH, that they are willing to pay per unit of gas, and the total transaction fee is computed as the product of gas consumed and the gas price. Since gas prices fluctuate with network demand, the same contract interaction can cost significantly different amounts in USD depending on current network conditions. 
For this reason, gas consumption is a more stable and implementation-specific metric than transaction fee, and is therefore used as the primary basis for comparing the two implementations in this work.
\Cref{tab:gas} summarizes the gas consumption of each contract method for both implementations.
To ensure a fair cost comparison, all USD values are recomputed using a unified gas price of 
$0.14754$ gwei and an ETH price of $2311.80$ USD/ETH.

\begin{table}[h]
\centering
\caption{Gas consumption comparison between Poseidon-based and SHA-256-based HTLC implementations.}
\label{tab:gas}
\begin{tabular}{lrrrr}
\toprule
\textbf{Method} & \multicolumn{2}{c}{\textbf{Avg Gas}} & \multicolumn{2}{c}{\textbf{Avg Cost (USD)}} \\
\cmidrule(lr){2-3} \cmidrule(lr){4-5}
                & Poseidon & SHA-256 & Poseidon & SHA-256 \\
\midrule
\texttt{createLock}        & 141,133 & 142,473 & 0.0481 & 0.0486 \\
\texttt{withdraw}          &  75,230 &  86,896 & 0.0256 & 0.0296 \\
\texttt{refund}            &  57,168 &  61,636 & 0.0194 & 0.0210 \\
\midrule
\textbf{Deployment}        & 721,122 & 725,895 & 0.2459 & 0.2476 \\
\bottomrule
\end{tabular}
\end{table}

The results reveal that the Poseidon-based implementation consumes slightly less gas than the SHA-256 counterpart for all transactional methods. Specifically, \texttt{withdraw}, the most cryptographically intensive operation, as it involves on-chain hash verification, consumes 75,230 gas under Poseidon compared to 86,896 gas under SHA-256, a reduction of approximately 13.4\%. Similarly, \texttt{createLock} and \texttt{refund} show marginal gas savings of 0.9\% and 7.2\%, respectively. These savings are consistent with Poseidon's design goal of being more efficient than traditional hash functions within arithmetic-circuit-friendly execution environments. The deployment cost of the Poseidon-based implementation is 721,122 gas compared to 725,895 gas for the SHA-256 baseline, a marginal difference of approximately 0.7\% in favor of the Poseidon-based implementation. This slight reduction is attributable to the struct packing and 
storage layout optimizations introduced in the contract, which reduced the deployed bytecode size despite the additional overhead of linking the \texttt{PoseidonT2} library at deployment time.

It is important to note that the USD costs reported above are computed under fixed network conditions and serve as a reference point only. In practice, transaction fees on Ethereum are a function of two volatile quantities: the gas price, which fluctuates with network load and congestion, and the ETH/USD exchange rate. During periods of high network activity, gas prices can increase by several orders of magnitude, making even gas-efficient contracts significantly more expensive to interact with. Conversely, the gas consumption figures reported in \Cref{tab:gas} are deterministic and independent of market conditions, providing a stable basis for comparing the two implementations.

\paragraph{Execution Latency on Local Network}
\Cref{tab:latency-local} reports the average execution latency of the two primary contract interactions measured over 20 runs on a local Hardhat network.

\begin{table}[h]
\centering
\caption{Execution latency on local Hardhat network (20 runs).}
\label{tab:latency-local}
\begin{tabular}{lrrrr}
\toprule
\textbf{Method} & \multicolumn{2}{c}{\textbf{Avg Latency (ms)}} & \multicolumn{2}{c}{\textbf{Std Dev (ms)}}\\
\cmidrule(lr){2-3} \cmidrule(lr){4-5}
                & Poseidon & SHA-256 & Poseidon & SHA-256 \\
\midrule
\texttt{createLock} & 80.85 & 82.80 & 9.69 & 10.33 \\
\texttt{withdraw}   & 91.25 & 83.05 & 8.31 &  9.21 \\
\texttt{refund}     & 79.50 & 82.30 & 7.34 &  8.93 \\
\bottomrule
\end{tabular}
\end{table}

The Poseidon-based implementation exhibits comparable latency to the SHA-256 baseline on the local network, with results varying by method. \texttt{createLock} averages 80.85 ms compared to 82.80 ms under SHA-256, and \texttt{refund} averages 79.50 ms compared to 82.30 ms, both marginally faster than the baseline. \texttt{withdraw} is the only method where Poseidon exhibits slightly higher latency, averaging 91.25 ms compared to 83.05 ms under SHA-256, which is attributable to the additional computational overhead of the Poseidon permutation during hash verification and the external library call to \texttt{PoseidonT2}. Notably, the standard deviations of the Poseidon implementation are lower than those of the SHA-256 baseline across all methods, suggesting more consistent execution behavior. Overall, these results indicate that the overhead of adopting Poseidon over SHA-256 is negligible in a local execution environment, with the two implementations performing 
at comparable levels across all three contract methods. It should be noted that latency measurements on a local Hardhat network reflect only the computational cost of transaction execution, as block times are instantaneous and there is no network propagation delay, making this an accurate isolated measure of the hash function overhead itself.

\paragraph{Execution Latency on Sepolia Testnet}
To obtain timing results reflective of real-world network conditions, the benchmark was additionally executed on the Sepolia Ethereum testnet. \Cref{tab:latency-sepolia} presents the results.

\begin{table}[h]
\centering
\caption{Execution latency on Sepolia testnet (20 runs).}
\label{tab:latency-sepolia}
\begin{tabular}{lrr}
\toprule
\textbf{Method} & \textbf{Avg Latency (ms)} & \textbf{Std Dev (ms)} \\
\midrule
\texttt{createLock} & 14,300.95 & 3,429.41 \\
\texttt{withdraw}   & 10,675.15 & 3,044.74 \\
\texttt{refund}     & 16,533.10 & 5,449.27 \\
\bottomrule
\end{tabular}
\end{table}

The Sepolia results are markedly different from the local measurements, with average latencies of 14.3 seconds for \texttt{createLock}, 10.7 seconds for \texttt{withdraw}, and 16.5 seconds for \texttt{refund}. This substantial increase is primarily explained by Ethereum's block time: unlike the local Hardhat network, where transactions are mined instantly, Sepolia operates with a block time of approximately 12 seconds under the proof-of-stake consensus mechanism. A submitted transaction must wait to be included in the next block, meaning the observed latency is dominated by this waiting period rather than by the computational cost of executing the contract logic itself.

The high standard deviations observed across all three methods further reflect the non-deterministic nature of block inclusion on a real network. Depending on when a transaction is submitted relative to the current block, the waiting time can vary between near-zero and a full block interval. 
Moreover, these figures represent average behavior under moderate network conditions, during periods of high congestion, latencies can increase substantially beyond what is reported here, as transactions offering lower gas prices may be deprioritized by block proposers in favor of more profitable ones. These results confirm that for time-sensitive applications built on top of this HTLC, the minimum lock duration parameter \texttt{minLockDuration} must be set with sufficient margin to account for realistic block inclusion delays and potential network congestion on the target network.

\section{Conclusion}
\label{sec:conclusion}

In this paper, we proposed a fairly compensated protocol for distributed information retrieval and augmentation in autonomous agent networks. The protocol addresses the fundamental evaluation-disclosure paradox by enabling retrieval agents to evaluate and rank candidate documents without learning their plaintext contents, while simultaneously guaranteeing that data providers are compensated only when valid information is successfully delivered.
To achieve this, we orchestrate threshold homomorphic encryption, zero-knowledge proofs, and blockchain-based hash time-locked contracts into a unified multi-phase framework, which provides confidentiality, correctness, integrity, and fairness against malicious adversaries.
As autonomous agents increasingly depend on distributed external information sources, we believe such mechanisms will become essential for enabling trustworthy and economically sustainable data marketplaces.

\bibliography{refs}
\bibliographystyle{plain}


\end{document}